# Time-Synchronized Full System State Estimation Considering Practical Implementation Challenges

Antos Cheeramban Varghese, *Student Member, IEEE*, Hritik Shah, *Student Member, IEEE*, Behrouz Azimian, *Student Member, IEEE*, Anamitra Pal, *Senior Member, IEEE*, and Evangelos Farantatos, *Senior Member, IEEE*.

*Abstract*—As the phasor measurement unit (PMU) placement problem involves a cost-benefit trade-off, more PMUs get placed on the higher voltage buses. However, this causes many of the lower voltage levels of the bulk power system to not be observed by PMUs. This lack of visibility then makes time-synchronized state estimation of the *full system* a challenging problem. We propose a Deep Neural network-based State Estimator (DeNSE) to overcome this problem. The DeNSE employs a Bayesian framework to *indirectly* combine inferences drawn from slow timescale but widespread supervisory control and data acquisition (SCADA) data with fast timescale but select PMU data to attain sub-second situational awareness of the entire system. The practical utility of the proposed approach is demonstrated by considering topology changes, non-Gaussian measurement noise, and bad data detection and correction. The results obtained using the IEEE 118-bus system show the superiority of the DeNSE over a purely SCADA state estimator and a PMU-only linear state estimator from a techno-economic viability perspective. Lastly, scalability of the DeNSE is proven by estimating the states of a large and realistic 2000-bus Synthetic Texas system.

## I. Introduction

Power utilities attain situational awareness of their transmission system through the process of state estimation. Particularly, state estimation provides the inputs for performing real-time contingency analysis, optimal power flow, and even network expansion planning [1]. Traditionally, state estimation was done using the supervisory control and data acquisition (SCADA) system. With the introduction of phasor measurement units (PMUs), SCADA-PMU hybrid state estimators as well as PMU-only linear state estimators were proposed. Recently, it has become necessary to perform state estimation at higher speeds (< 0.1s) to understand the impacts of rapid fluctuations in outputs of converter-interfaced resources on the security of the bulk power system (BPS) [2]. However, purely SCADA state estimators and SCADA-PMU hybrid state estimators are not able to provide sub-second situational awareness, while PMU-only linear state estimators require PMUs to be optimally placed throughout the system. This paper proposes *a novel Bayesian framework for transmission system state estimation (TSSE) that indirectly combines inferences drawn from slow timescale but widespread SCADA data with fast timescale but select PMU data to attain high-speed (sub-second) situational awareness of the entire BPS (69kV and above)*.

### A. Motivations

Due to the asynchronous nature of their inputs, purely SCADA state estimators suffer from problems such as non-linearity, divergence, and low accuracy [3]. These problems will exacerbate with increase in the penetration of renewable generation. Hybrid state estimators directly combine data from SCADA and PMU systems [4]-[6]. Hence, they suffer from problems such as imperfect synchronization and time-skew errors [7]. Moreover, strategies proposed to overcome some of these problems [8]-[10] are compute intensive, which makes hybrid state estimators operate at slower timescales [11]. PMU-only linear state estimators provide time-synchronized outputs and are extremely fast, but they require the system to be fully observed by PMUs [12]. The unobservability issue associated with PMU-only linear state estimation (LSE) is typically relegated to solving an optimal PMU placement (OPP) problem [13]-[18]. However, many OPP formulations minimize the number of PMUs, which does not translate to minimizing the cost of PMU deployment [19]. This happens because the main drivers of PMU installation costs are communication, security, and labor [20], and they increase with the number of substations that are upgraded for PMU placement. Now, as the highest voltage buses/substations are the backbone of the BPS, and these buses are fewer in number, they become the natural choice for placing the PMUs. Conversely, placing PMUs at lower voltage levels do not yield as much benefits. This cost-benefit trade-off and law of diminishing returns prevents the lower voltage levels from being fully observed by PMUs.

We have investigated the reality of the PMU-unobservability problem by collecting data from two U.S. power utilities as shown below in Table I and Table II, respectively. Table I shows the PMU coverage of a U.S. power utility located in the Eastern Interconnection (EI). This power utility has more than 1,400 buses, but only 129 of them are equipped with PMUs. Moreover, as the voltage levels decrease, there is a sharp drop in the number of buses with PMUs to the total number of buses at that voltage level. This confirms that PMUs are mostly placed on higher voltage buses. Lastly, from the last column of Table I it can be realized that none of the voltage levels are fully observed by PMUs implying that PMU-only LSE cannot be performed at any voltage level of this power utility.

Manuscript received: August 02, 2023, Accepted: February 20, 2024, Date of cross-check: xx, Date of online publication: xx

This work was supported in part by the U.S. Department of Energy under grant DE-EE0009355, the National Science Foundation (NSF) under grant ECCS-2145063, and the Electric Power Research Institute (EPRI) under grant 10013085. The views expressed herein do not necessarily represent the views of the U.S. Department of Energy or the U.S. Government.

A.C. Varghese, H. Shah, B. Azimian, and A. Pal are with the School of Electrical, Computer, and Energy Engineering of Arizona State University, Tempe, AZ 85281 USA, (email: antos.varghese@asu.edu; hshah59@asu.edu; bazimian@asu.edu; Anamitra.pal@asu.edu).

E. Farantatos is with EPRI, 3420 Hillview Ave, Palo Alto, CA, 94304 (email: efarantatos@epri.com).

A. C. Varghese and H. Shah are co-first authors.

DOI:

Table II shows the PMU coverage of a U.S. power utility that is part of the Western Electricity Coordinating Council (WECC). A key difference in this table in comparison to Table I is that the third column denotes the number of PMU devices, and not the number of PMU equipped buses. Furthermore, it can be realized from Table II that despite having a large number of PMUs at different voltage levels, none of the levels are completely observed by them. This happens because (a) PMUs serve other functions than state estimation [13], and (b) the cost of adding more devices at one substation is incremental [21]-[22], and so power utilities add more PMUs to the same location even if they do not aid state estimation. Thus, *high-speed time-synchronized state estimation for a transmission system that is only locally observed by PMUs is a challenging practical problem*. In the rest of the paper, the terms locally observable and (PMU)-unobservable will be used interchangeably.

TABLE I
PMU COVERAGE OF A U.S. POWER UTILITY IN THE EI

| Voltage-level | #Buses | #PMU equipped buses | % observed |
|---|---|---|---|
| 500kV | 52 | 28 | 79 |
| 230kV | 15 | 5 | 53 |
| 161kV | 1185 | 92 | 27 |
| 115kV | 42 | 2 | 10 |
| 69kV | 144 | 2 | 3 |

TABLE II
PMU COVERAGE OF A U.S. POWER UTILITY IN THE WECC

| Voltage-level | #Buses | #PMUs | % observed |
|---|---|---|---|
| 500kV | 18 | 53 | 90 |
| 230kV | 47 | 89 | 80 |
| 115kV | 30 | 23 | 30 |
| 69kV | 258 | 207 | 50 |

To counteract impact of unobservability on state estimation, pseudo-measurements obtained by interpolated observations or forecasts obtained using historical data, can be used. However, as demonstrated in [23], such approaches do not ensure quality of the estimates. Recently, machine learning (ML) has been used to address the observability issues w.r.t. high-speed state estimation [24]-[26]. Ref. [24] proposed a Bayesian state estimator using deep neural networks (DNNs), but it was tailored for distribution systems. An ML-based state estimator for incompletely observed transmission systems was created in [25]. A state estimator with two DNNs (one for observable part and the other for unobservable part of the system) was proposed in [26]. However, [25], [26] did not consider the practicality of sensor placement when creating the ML-based state estimators.

*B. Novel contributions of the paper*

Motivated by the knowledge gaps outlined in the previous sub-section, we propose a Deep Neural network-based State Estimator (DeNSE) that estimates all the transmission system voltages in a time-synchronized manner from PMUs that are only placed at the highest voltage buses of the system. By performing TSSE using very few PMUs, the DeNSE also circumvents the need for a massive supporting communication infrastructure [27]. Apart from the unobservability issue, this paper addresses four other practical challenges that exist w.r.t. high-speed time-synchronized TSSE as summarized below.

The first is the *scalability* of the state estimation technique. The classical LSE formulation involves a matrix inversion step, whose computational complexity is $O(n^{2.3727})$ [28]. As such, the time consumption of this implementation increases quadratically w.r.t. the number of states. Conversely, during online implementation, the forward propagation of a neural network (NN) only involves multiplication and addition operations, whose complexity ($O(n \log n)$) is much lower [29]. The second is the *presence of non-Gaussian noise* in PMU measurements [30]-[33]. The LSE formulation is the solution to the maximum likelihood estimation (MLE) problem under Gaussian noise environments. This means that its performance can deteriorate in presence of non-Gaussian noise. However, a NN-based framework, such as the DeNSE, does not have such a limitation. The third is *high-speed bad data detection and correction (BDDC)* [34]. Dearth of measurements makes this challenge particularly acute for the problem being solved here. To address this challenge, a robust BDDC algorithm based on a combination of the Wald test [35] and an extreme scenario filter, is developed. The fourth is topology changes. This is a major concern for NN-based state estimators because it results in the training and testing environments (of the NNs) to differ, which can then deteriorate their performance. This challenge is tackled by combining DeNSE with topology processor outputs and Transfer learning [24], [36].

In summary, this paper advances the state-of-the-art for time-synchronized state estimation in transmission systems by making the following salient contributions:
1. A high-speed time-synchronized state estimator called the DeNSE is developed for the BPS that overcomes the need to fully observe the system by PMUs.
2. A robust BDDC methodology is created that ensures performance of the DeNSE under diverse types of bad data and loading conditions.
3. The ability of the DeNSE to tackle topology changes and non-Gaussian measurement noise is demonstrated.

We also provide a logical explanation along with a numerical example in Appendix A to illustrate how DeNSE can perform state estimation for unobservable power systems.

## II. PROPOSED FORMULATION

*A. Bayesian-approach to TSSE*

PMU-only LSE solves a variant of the MLE problem, with the most common being the least squares formulation. However, the least squares solution requires the system of equations to have full rank, which translates to the well-known constraint of complete system observability by PMUs. One way to circumvent this constraint is to reformulate the TSSE problem within a Bayesian framework where the states, $x$, and the PMU measurements, $z$, are treated as random variables. Then, the following minimum mean squared error (MMSE) estimator can be formulated:

$$\min_{\hat{x}(\cdot)} \mathbb{E}(\|x - \hat{x}(z)\|^2) \Rightarrow \hat{x}^*(z) = \mathbb{E}(x|z) \quad (1)$$

Eq. (1) directly minimizes the *estimation error* without the knowledge of the physical model of the system. Note that in the classical LSE formulation, $z = Hx + e$, the *modeling error* is minimized, which is embedded in the measurement matrix, $H$. By avoiding the explicit need for $H$, the observability requirement is no longer necessary in the Bayesian framework. Furthermore, by directly minimizing the estimation error, no limitation (Gaussian/non-Gaussian) is imposed on the characteristics of the measurement noise, $e$.

However, there are two challenges in computing the expected conditional mean of (1). First, the conditional expectation, defined by $\mathbb{E}(x|z) = \int_{-\infty}^{+\infty} x p(x|z) dx$, requires the knowledge of the joint probability distribution function (PDF) between $x$ and $z$, denoted by $p(x, z)$. When the number of PMUs is scarce, $p(x, z)$ is unknown or impossible to specify, making direct computation of $\hat{x}^*(z)$ intractable. Second, even if the underlying joint PDF is known, finding a closed-form solution for (1) can be difficult. The DNN used in DeNSE overcomes these difficulties by providing an approximation of the conditional expectation of the MMSE estimator.

*B. Architecture of the DNN in the DeNSE*

The DNN has a feed-forward architecture with $m$ inputs and $n$ outputs, where $m$ corresponds to the measurements coming from PMUs and $n$ refers to the total number of states to be estimated (i.e., $z \in \mathbb{R}^m$ and $x \in \mathbb{R}^n$). Due to incomplete observability of the system by PMUs, $m \ll n$. The DNN has $h$ hidden layers, in which the input vector entering the $(i+1)^{th}$ layer is expressed in terms of the inputs from the $i^{th}$ layer as:

$$c_{i+1} = W_{i+1,i} d_i + b_{i+1} \quad (2)$$

where $c_{i+1}$ represents the input vector entering the $(i+1)^{th}$ layer, $W_{i+1,i}$ represents the weight between the $i^{th}$ and the $(i+1)^{th}$ layer, $d_i$ denotes the output of the $i^{th}$ layer, and $b_{i+1}$ represents the bias value of the $(i+1)^{th}$ layer. Next, $c_{i+1}$ is passed through an activation function, $a_{i+1}$, to yield $d_{i+1}$:

$$d_{i+1} = a_{i+1}(c_{i+1}) \quad (3)$$

This propagation continues through all the hidden layers and the resulting value is obtained at the output layer. The loss function compares the estimated output and corresponding true output. The error between them is represented by:

$$\varepsilon_j = \zeta(o_j, \hat{o}_j) \quad (4)$$

where $\varepsilon_j$ denotes the error, $o_j$ denotes the true value of the output and $\hat{o}_j$ denotes the estimated value of the output by the DNN in the current epoch; $\zeta$ is an appropriate loss function that indicates how well the DNN has been trained. To improve the training accuracy, $\zeta$ is minimized by optimally tuning the weights and biases through a process called backpropagation. The process is repeated until the loss becomes acceptable.

*C. Creation of Training Database*

A unique feature of the DeNSE framework that sets it apart from other ML-based state estimators (such as [37]) is that it *does not use* the slow timescale measurements to directly train the DNN. Instead, the discrete power injection measurements from the SCADA system are first converted into continuous functions by fitting an appropriate distribution to them. Then, independent Monte Carlo (MC) sampling is employed to randomly sample points from the distribution to feed as inputs to a power flow solver. The power flow is solved a large number of times, providing voltage and current phasor values across all system buses under various operating conditions. Then, for training, we use voltage and current phasors (with added noise) of buses which are equipped with PMUs as inputs to the DNN, while voltage phasors of all the buses are set as outputs of the DNN. This process helps in capturing the uncertainty introduced by the load variations and makes the DNN aware of diverse loading conditions.

Training the DNN by using the above-mentioned process of indirectly combining inferences from SCADA and PMU data has two advantages: (i) *the problem of temporal differences and synchronization issues are completely circumvented*, and (ii) *any reasonable errors in the SCADA data do not impinge on the performance of the DeNSE*. The DeNSE *can* be impacted by bad PMU data and/or noise in the data since this data is the input to the trained DNN during online operation. Input data quality effects are investigated analytically in Section III.B, and experimentally in Sections IV.B, IV.E, and IV.F, respectively.

III. ENHANCEMENTS TO PROPOSED FORMULATION AND ONLINE IMPLEMENTATION

*A. Transfer learning to handle topology changes*

A DNN trained using the formulation proposed in Section II will perform fast and accurate time-synchronized state estimation for PMU-unobservable BPS during real-time operation as long as the topology does not change. However, if the topology used for training and testing differs, then the joint PDF between the measurements and the states will change; this can deteriorate the performance of the DeNSE. A possible alternative is to train the DNN from scratch for the new topology. However, it will take a very long time to do so. Instead, we use Transfer learning to update the DNN of the DeNSE when topology changes. Transfer learning refers to utilizing models learned from an old problem and leveraging them for a new problem, in order to maintain learning performance and accuracy. In the context of TSSE, Transfer learning is particularly useful because when a topology change occurs, the mapping between measurements and states of only a small portion of the system gets altered. This implies that the re-learning will be localized.

We employ inductive transfer learning [38], to induce knowledge transfer from the old (base) topology to the new (current) topology. Four approaches have been proposed for implementing inductive transfer learning: feature-representation transfer, instance transfer, relational-knowledge transfer, and parameter transfer. We use parameter transfer to update the parameters of the DNN when topology changes. Two well-known parameter transfer methods are parameter-sharing and fine-tuning. Parameter-sharing assumes that the parameters are highly transferable due to which the parameters in the source domain (old topology) can be directly copied to the target domain (new topology), where they are kept "frozen". Fine-tuning assumes that the parameters in the source domain are useful, but they must be trained with limited target domain data to better adapt to the target domain [39]. Since there is no guarantee that the parameters of the DNN will be highly transferable for different topologies, *fine-tuning* is used in this paper for Transfer learning.

To determine when Transfer learning via fine-tuning should be implemented, we make use of the topology processor of the BPS. After updating the DNN, the new topology is designated as the base topology, to make it consistent with the DeNSE. The overall implementation is shown in Figure 1.

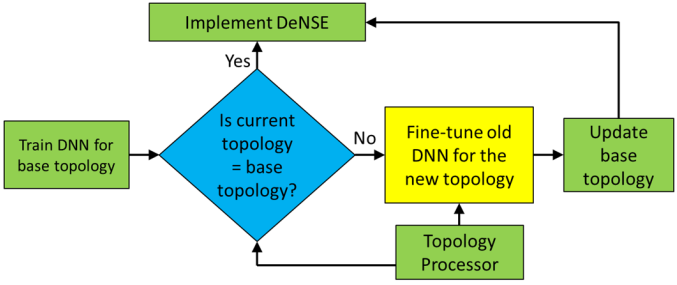

Figure 1: Implementation of Transfer learning to handle topology changes

### B. Robust bad data detection and correction (BDDC)

During online implementation, streaming PMU data will be fed as inputs to the proposed DeNSE framework. However, PMU data obtained from the field often suffers from bad data in the form of data dropouts and outliers [40]. This is different from measurement noise since bad data have very different amplitudes compared to normal noisy data. To prevent such data from impacting the performance of the DeNSE, a robust BDDC methodology capable of operating at PMU timescales (≤33 ms) is devised as a precursor to this state estimator.

*1) BDDC using Wald test*

A technique to detect bad data before it can enter an ML-based state estimator was proposed in [23]. The technique relied on the Wald test [35] to flag incoming measurements as bad. To apply this test, two hypotheses must be defined first: (a) $H_0$: models the measurement without bad data and has a distribution with mean, $\mu_0$, and variance, $\sigma_0^2$, both of which are learned during training. (b) $H_1$: models the measurement with bad data, because of which its mean and variance is very different from that of $H_0$. Mathematically, the Wald test can be expressed as:

$$\left|\frac{z - \mu_0}{\sigma_0}\right| \gtreqless_{H_0}^{H_1} = Q^{-1}\left(\frac{\alpha}{2}\right) \quad (5)$$

In (6), $Q(y) = \frac{1}{\sqrt{2\pi}}\int_y^\infty \exp\left(\frac{-u^2}{2}\right)du$, is the tail of the distribution and $\alpha$ is a tunable parameter that specifies the false positive limit. Essentially, the Wald test makes use of the fact that DNN training is done using good quality data. Hence, once the limits of good quality data become known during training, any testing data that lies outside that limit can be termed as bad. This Wald test-based bad data detection method developed in [23] was found to be compatible with the high-speed requirements of the DeNSE. However, [23] corrected the identified bad data by simply replacing it with its mean value from the training database. The methodology for correcting the bad data is different, as explained below.

Since the Wald test is applied independently and simultaneously to all the $m$ input features of a given sample of the testing dataset, it is unlikely that all the features will be bad at the same time. For a given testing dataset sample, $\mathbf{z}_{\text{sample}}^{\text{test}}$, let the set of indices that correspond to features flagged as bad by the Wald test be called ibfs. Then, if iafs denotes the set of indices corresponding to all the features of $\mathbf{z}_{\text{sample}}^{\text{test}}$, the difference of these two sets gives the set of indices corresponding to the good features of $\mathbf{z}_{\text{sample}}^{\text{test}}$; let this set be denoted by igfs. Now, igfs can be used to find that OC in the training database, $\mathbf{Y}^{\text{train}}$, that most closely resembles the OC captured by $\mathbf{z}_{\text{sample}}^{\text{test}}$. Once that OC (called nearest OC (NOC)) is found, its entries corresponding to ibfs should replace the flagged features of $\mathbf{z}_{\text{sample}}^{\text{test}}$. The overall methodology is depicted in **Algorithm I** and is performed for every sample of the testing dataset. The superiority of the proposed bad data correction methodology over the one where it is replaced with mean values is demonstrated in Section IV.E.

| **Algorithm I:** Novel bad data correction methodology using nearest OC (NOC) in training dataset |
|---|
| **Input:** $\mathbf{z}_{\text{sample}}^{\text{test}}$, $\mathbf{Y}^{\text{train}}$ |
| **Output:** The corrected testing dataset sample, $\mathbf{z}_{\text{sample\_crct}}^{\text{test}}$ |
| 1. Create array of indices, iafs, from $\mathbf{z}_{\text{sample}}^{\text{test}}$, and set $\mathbf{z}_{\text{sample\_crct}}^{\text{test}} = \mathbf{z}_{\text{sample}}^{\text{test}}$ |
| 2. Conduct Wald test on $\mathbf{z}_{\text{sample}}^{\text{test}}$ and flag the indices of bad data to create ibfs |
| 3. {igfs} = {iafs} − {ibfs} |
| 4. $k^* = \arg\min_k \|\mathbf{Y}^{\text{train}}[k, \text{igfs}] - \mathbf{z}_{\text{sample}}^{\text{test}}[\text{igfs}]\|$ |
| 5. $\mathbf{z}_{\text{sample\_crct}}^{\text{test}}[\text{ibfs}] = \mathbf{Y}^{\text{train}}[k^*, \text{ibfs}]$ |

*2) Differentiating between bad data and extreme scenarios*

The Wald test is very sensitive to the choice of $\alpha$. A very small value of $\alpha$ may result in bad data being treated as good data, while a large value may result in an extreme scenario data being treated as bad data. This can happen because by definition extreme scenarios are those OCs that are unlikely to occur normally. In the worst-case, data corresponding to an extreme scenario will get flagged as bad data and be replaced by normal data from the training database, making the DeNSE produce an incorrect picture of the operating state of the system. We combine our knowledge of how PMUs are placed in a power system with how extreme OCs actually manifest, to design an *extreme scenario filter* that prevents this problem.

Furthermore, if PMUs are placed only at the highest voltage buses (which is the premise of this paper), they will be automatically (electrically) close to each other even for PMU-unobservable BPS. This is because the highest voltage buses are connected to each other by the highest voltage lines. Thus, when an extreme scenario manifests, measurements of multiple PMUs will be simultaneously impacted. Conversely, bad data occurs randomly in both space and time. This realization leads us to propose the following logic for the design of the extreme scenario filter: *If one or more features of the testing data sample are simultaneously identified as bad by the Wald test for* p *different PMUs, each of which are within* p *hops of each other, then the data sample corresponds to an extreme OC and should not be treated as bad data*. This logic is implemented in the manner shown in **Algorithm II**.

Note that in **Algorithm II**, $p$ indicates the severity of the extreme scenario; higher the value of $p$, a greater number of hops to be considered. Lastly, the extreme scenario filter is combined with the proposed BDDC methodology in the following way: whenever the filter gets activated, the results of the Wald test are suppressed (i.e., no data correction occurs), and the raw PMU measurements are fed as inputs to the trained DNN of the DeNSE. The usefulness of the extreme scenario filter in making the working of DeNSE more realistic is demonstrated in Section IV.F.

| **Algorithm II:** Extreme scenario filter implementation |
|---|
| **Input:** Features flagged as bad by Wald test, ibfs |
| **Output:** Features passing extreme scenario filter, $ibfs_{ESF}$ |
| 1. $ESF_{ini}$ = PMU locations corresponding to ibfs |
| 2. p = Length ($ESF_{ini}$) |
| 3. $ibfs_{ESF}$ = ibfs |
| 4. $ESF_p$ = List of subsets of $ESF_{ini}$ with p elements |
| 5. For (k = 1: Length($ESF_p$)): |
|    a. If (every element of $ESF_p$ [k] is within p hops of each other): |
|      i. $Feat_{ESF}$ = List of all features corresponding to $ESF_p$[k] |
|      ii. $\{ibfs_{ESF}\} = \{ibfs_{ESF}\} - \{Feat_{ESF}\}$ |
|    b. End If |
| 6. End For |
| 7. p = p − 1 |
| 8. If (($ibfs_{ESF} \neq ibfs$) or (p < 2)): |
| 9.   End |
| 10. Else Go to Step 3 |

### C. Implementation of DeNSE

Figure 2 shows the overall framework for the proposed DeNSE. It has an offline learning phase and an online implementation phase. In the offline phase, appropriate distributions are fitted to historical SCADA data using Kernel density estimation (KDE). MC sampling is done from the fitted distributions and set as inputs to a power flow solver to generate training data for the DNN. The voltage and current phasors corresponding to actual PMU locations are used to train the DNN while all the voltage phasors (states) are set as outputs of the DNN. The DNN approximates the conditional expectation shown in (1). While (1) holds true for measurements in the polar or rectangular form, the DeNSE is implemented in polar form since (a) PMUs report in that form, and (b) a DNN is capable of approximating non-linear functions effectively (note that the relation between measurements and states in polar form is non-linear). Once the optimized DNN parameters are found, the DNN training is complete. In the online phase, streaming PMU data is passed through the Wald test and a data preprocessing block (based on Section III.B), and the resulting samples are sent to the trained DNN to produce the state estimates.

## IV. RESULTS AND DISCUSSION

### A. State estimation results for IEEE 118-bus system

The effectiveness of the DeNSE is first illustrated using the IEEE 118-bus system. Each bus of this system is mapped to a bus in the 2000-bus Synthetic Texas system [41]-[42] of similar mean power rating. This is done because the Texas system has one-year of SCADA data publicly available, and this mapping helps in obtaining realistic variations in the active and reactive powers for every bus of the 118-bus system. Next, the power injection distributions are found using KDE. After picking samples independently from the distributions, a power flow is solved to create the training, validation, and testing data.

It is assumed that PMUs are only placed on the highest voltage buses of this system, namely, 8, 9, 10, 26, 30, 38, 63, 64, 65, 68, and 81. PMUs located at these 11 buses measure the voltage of the corresponding bus as well as the currents flowing in the lines emanating from that bus. The 41 PMU measurements (= 11 bus voltage phasors + 30 branch current phasors) are the inputs to the DNN. The outputs of the DNN are the 118 voltage magnitudes and angles of this system.

The training and testing of the DNN is carried out using Keras with TensorFlow as the backend library in Python [43]. Training a DNN involves finding hyperparameter values that give desired performance. The basic hyperparameters of a DNN are the number of hidden layers, the number of neurons per layer, and the activation function. The activation function used in the hidden layers is rectified linear unit (ReLU), while a linear function is used in the output layer. To overcome the problem of internal covariate shift, batch normalization is employed. Dropout regularization is used to prevent DNN overfitting. The mean squared error (MSE) loss function is used to calculate the error between the predicted and the true states. During back-propagation, the Adam optimizer is used to update the weights of the DNN. Table III summarizes the optimal values of the hyperparameters obtained for the DeNSE for the 118-bus system; hyperparameter tuning was done using the ML platform WANDB [44]. All simulations were performed on a computer with 256 GB RAM, Intel Xeon 6246R CPU @3.40GHz, Nvidia Quadro RTX 5000 16 GB GPU. All codes for this paper can be accessed using the GitHub link provided in Appendix B.

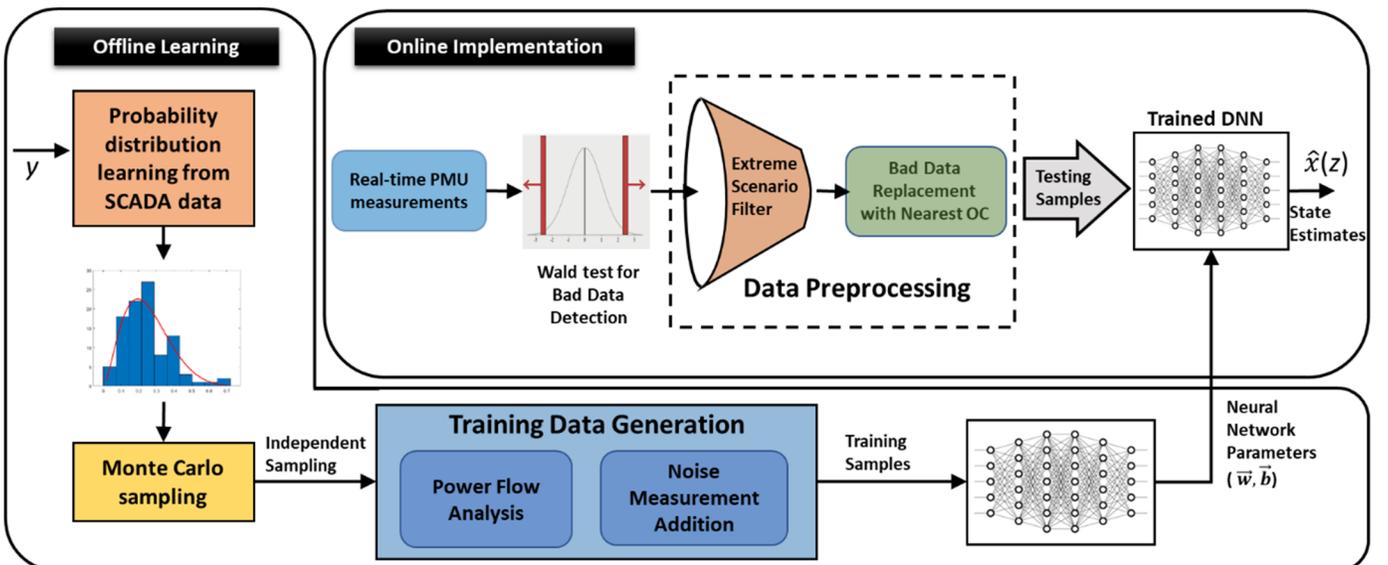

Figure 2: Proposed Bayesian framework for DeNSE

TABLE III
HYPERPARAMETERS OF DENSE FOR IEEE 118-BUS SYSTEM

| Hyperparameter | Value |
|---|---|
| Number of Hidden Layers | 4 |
| Number of Neurons per Hidden Layer | 500 |
| Activation Functions | ReLU (Hidden Layers) Linear (Output Layer) |
| Loss Function | Mean Squared Error |
| Optimizer | Adam |
| Batch Size | 128 |
| Learning Rate | 0.0207 |
| Number of Epochs | 2,000 |
| Early Stopping | Patience = 10 |
| Dropout | 30% |
| Dataset size | |
| Training | 7,500 |
| Validation | 2,500 |
| Testing | 4,000 |
| Total | 14,000 |

Figure 3 shows the average error in the voltage angle and magnitude estimates as a function of the distance from the buses where the PMUs are placed. The error metrics used were mean absolute percentage error (MAPE) for magnitudes and mean absolute error (MAE) for angles. The distance is expressed in terms of hops from the bus where the PMU is placed; i.e., a hop of zero corresponds to the 11 highest voltage buses of this system. It is clear from the figure that in comparison to conventional approaches (such as LSE) that are limited to hops of zero and one (i.e., the observable regions of the system), the DeNSE is able to give reasonable state estimates even for buses that are six or seven hops away.

*B. Analyzing impact of measurement noise*

The plots shown in Figure 3 were obtained under 1% total vector error (TVE) [45] Gaussian noise environments. Now, it is important to analyze the impact that different types of noises will have on the performance of a data-driven state estimator such as the DeNSE. It has recently been shown that PMU noises can have non-Gaussian characteristics [30]-[31]. Keeping this in mind, three types of noise characteristics are considered in this study – Gaussian noise, Gaussian mixture model (GMM) noise [32], and Laplacian noise [33]. The Gaussian noise had zero mean, and standard deviation of 0.0033% in magnitude and 0.0029 rad in angle. The GMM noise had two-components having mean, standard deviation, and weight vectors as [0, 0.005]%, [0.0015, 0.0015]%, and [0.3, 0.7], in magnitude, and [0, 0.0043] rad, [0.0014, 0.0014] rad, and [0.3, 0.7], in angle, respectively. The Laplacian noise had a location and scale of 0.001% and 0.0015% in magnitude, and 0.0009 rad and 0.0013 rad, respectively, in angle. The above-mentioned noise parameters corresponded to a TVE of 1%. The results obtained using the DeNSE in presence of these three noise models are shown in Table IV. From the table, it is observed that the DeNSE is robust enough to handle non-Gaussian measurement noise in an effective manner as there is only a very minor deterioration in performance as the noise models change.

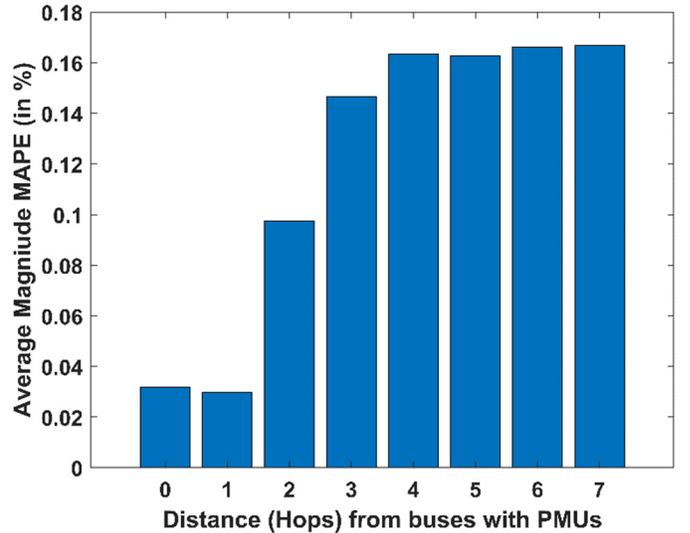

(a)

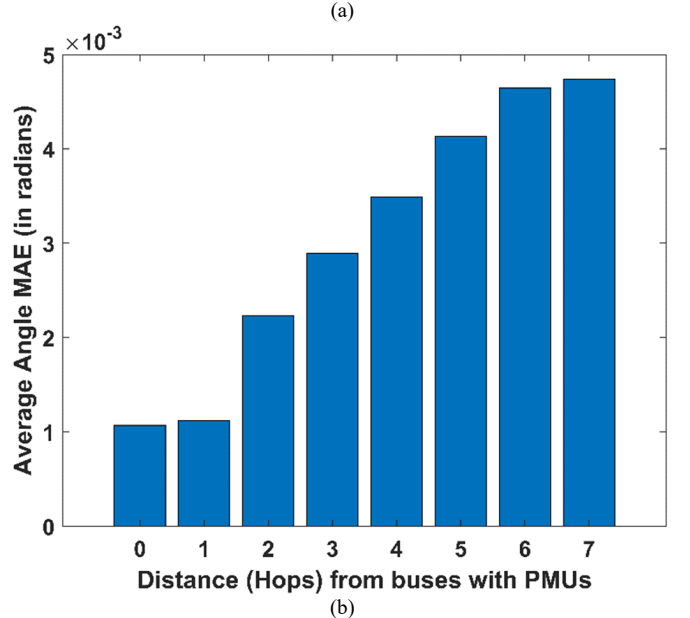

(b)

Figure 3: Performance evaluation of the DeNSE for the IEEE 118-bus system as a function of the distance from the PMU-placed buses

TABLE IV
PERFORMANCE OF DENSE UNDER DIFFERENT NOISE MODELS FOR IEEE 118-BUS SYSTEM

| Noise Model | Ave. Mag. MAPE (%) | Ave. Angle MAE (rad) |
|---|---|---|
| Gaussian | 0.1676 | 0.0042 |
| GMM | 0.1667 | 0.0047 |
| Laplacian | 0.1678 | 0.0049 |

*C. Comparing with other state estimators*

The performance of the DeNSE is now compared with two other state estimators, namely, a purely SCADA state estimator, and a PMU-only linear state estimator. For fairness of comparison, 1% TVE Gaussian noise was added to all the PMU measurements. The SCADA measurements comprise all sending-end active power flows and voltage magnitudes [46], corrupted by 10% additive Gaussian noise. The linear state estimator received PMU data from 32 buses that were identified from OPP studies [13]. Table V presents the average magnitude MAPE and angle MAE for all three state estimators. It is clear from the table that the purely SCADA-based state estimator has testxbxinferior performance compared to the DeNSE in terms of both magnitude and angle estimation. Although the PMU-only linear state estimator gives similar performance as the DeNSE, it requires almost three times the number of PMUs, that too placed optimally in the system. Thus, considering the practical implementation challenges associated with time-synchronized TSSE, the DeNSE results are optimal from a techno-economic viability perspective.

We also compared the performance of DeNSE with the NN-based state estimator developed in [26]. The results are shown in Table VI. Note that [26] had placed PMUs at 32 locations (compared to 11 in our case). However, these locations were not optimally selected, resulting in five buses being unobservable in [26] for the 118-bus system. From Table VI, the following inferences are drawn: (a) The proposed DeNSE has a higher root mean squared error (RMSE). This is due to the fact that the number of locations where PMUs are placed is almost one-third in our case. (b) The proposed DeNSE is more robust to noise. This is because with increase in noise amplitude (standard deviation of the noise), there is a two order of magnitude increase in the RMSE values of [26], whereas there is only a 15% increase in the RMSE values of the proposed DeNSE as the noise amplitude increases.

TABLE V
COMPARISON OF DeNSE WITH OTHER OPTIMIZATION-BASED STATE ESTIMATORS FOR IEEE 118-BUS SYSTEM

| Type | # PMU Locations | Ave. Mag. MAPE (%) | Ave. Angle MAE (rad) |
|---|---|---|---|
| Purely SCADA state estimator | - | 0.9816 | 0.0079 |
| PMU-only linear state estimator | 32* | 0.2709 | 0.0026 |
| DeNSE | 11 | 0.1676 | 0.0042 |

* Optimally placed to ensure complete system observability

TABLE VI
COMPARISON OF DeNSE WITH THE NN-BASED STATE ESTIMATOR DEVELOPED IN [26] FOR IEEE 118-BUS SYSTEM

| Noise amplitude (std. deviation of noise) | RMSE of [26] with PMUs at 32 buses* | RMSE of DeNSE with PMUs at 11 buses |
|---|---|---|
| 0 | $2.28 \times 10^{-6}$ | $6.29 \times 10^{-3}$ |
| 0.001 | $1.86 \times 10^{-5}$ | $6.60 \times 10^{-3}$ |
| 0.01 | $2.00 \times 10^{-4}$ | $6.70 \times 10^{-3}$ |
| 0.03 | $5.00 \times 10^{-4}$ | $6.94 \times 10^{-3}$ |
| 0.05 | $9.00 \times 10^{-4}$ | $7.22 \times 10^{-3}$ |

* Not optimally placed (five buses left unobserved)

### D. Investigating impact of topology changes

Next, we investigate the ability of Transfer learning in updating the DNN of DeNSE after a topology change takes place. A set of likely topologies were identified for the 118-bus system by removing one line at a time between any two buses of the system such that an island is not formed; 177 such topologies were identified. The training data for these likely topologies were saved in the database. When a topology change is detected by the topology processor in real-time, Transfer learning via fine-tuning is activated as described in Figure 1. The results obtained are as follows.

Let the base topology be denoted by $T_1$. By opening different lines, three new topologies were created from $T_1$. $T_2$ was created by opening the line between buses 75 and 77, neither of which have a PMU on them. $T_3$ was obtained when the line between buses 38 and 37 was removed; note that bus 38 has a PMU on it. $T_4$ was realized by opening the line between buses 26 and 30, both of which have a PMU on them. Figure 4 and Figure 5 show the changes in topology and their influence on TSSE with and without Transfer learning.

When Transfer learning is used to update the DNN, fine-tuning only takes 30 seconds of re-training time to give similar results for the new topologies as was obtained for the base topology (the heights of the grey and blue bars are similar). Note that if we had trained the DNN from scratch for every new topology, it would have taken 3 hours for every topology change, making the DeNSE inconsistent with the current state of the system for a much longer time-period. The reason why fine-tuning is so fast is because it only needs 2,000 samples and 90 epochs compared to 10,000 samples and 2,000 epochs that were needed to train the DNN from scratch (see Table III). Conversely, if the DNN trained for $T_1$ is used throughout, the performance of DeNSE degrades significantly (compare the heights of blue and orange bars in Figure 4 and Figure 5).

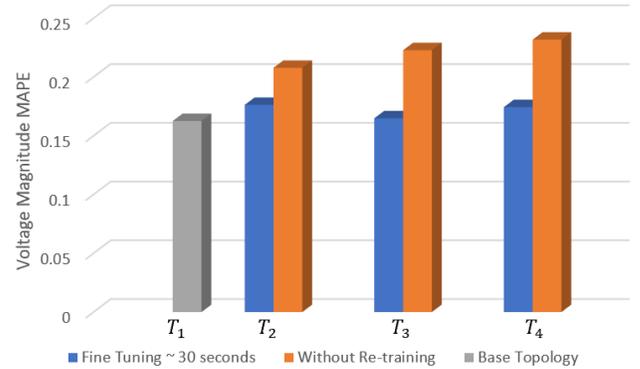

Figure 4: Efficacy of Transfer learning – average magnitude MAPE in %

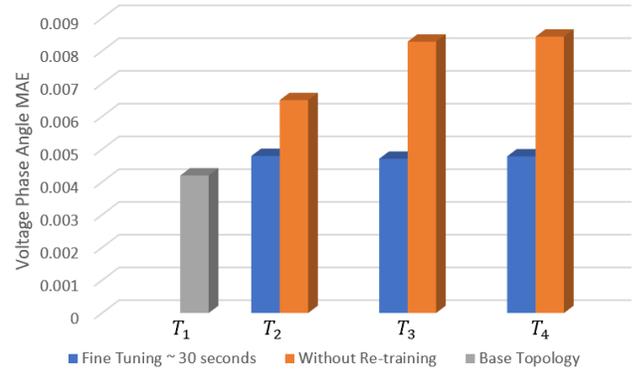

Figure 5: Efficacy of Transfer learning – average angle MAE in rad

It can also be observed from the two figures that the deterioration in estimation is more prominent for $T_3$ and $T_4$. This happened because the line that was opened for creating these two topologies had PMUs placed on one and both ends of the line, respectively. Due to this, when the line opened, the outputs of these PMUs became very different from what they were during the training of the DNN. This culminated in the considerable difference in the training and testing environments after the topology change occurred, causing increased deterioration in the performance of the trained DNN.

*E. Mitigating impact of bad data*

To investigate the performance of the proposed nearest OC (NOC)-based BDDC methodology, we simulated two different scenarios. In the first scenario, we increase the amount of testing samples that are bad, while fixing the severity of the bad data. To do this, the probability of bad data was randomly varied from $\eta = 0\%$ to $\eta = 50\%$ in steps of 10%, while the severity was kept at $\sigma = 3\sigma_0$, where $\sigma_0$ denotes the standard deviation of good quality data computed from the training dataset. The value of $\alpha$ was set to 0.05 to ensure that the false alarm (false positive) probability does not exceed 5%. The results obtained when the proposed methodology is compared with a case where the bad data is not replaced and a case where the bad data is replaced with the mean value from the training dataset (as done in [23]), are shown in Figure 6. It is clear from the figure that in the absence of BDDC, the results become progressively worse as the amount of bad data increases (red dotted line). Moreover, it can be observed that the bad data correction based on the NOC consistently outperformed the bad data correction based on the mean value for both magnitude and angle estimation (the green dotted line always lay below the blue dotted line).

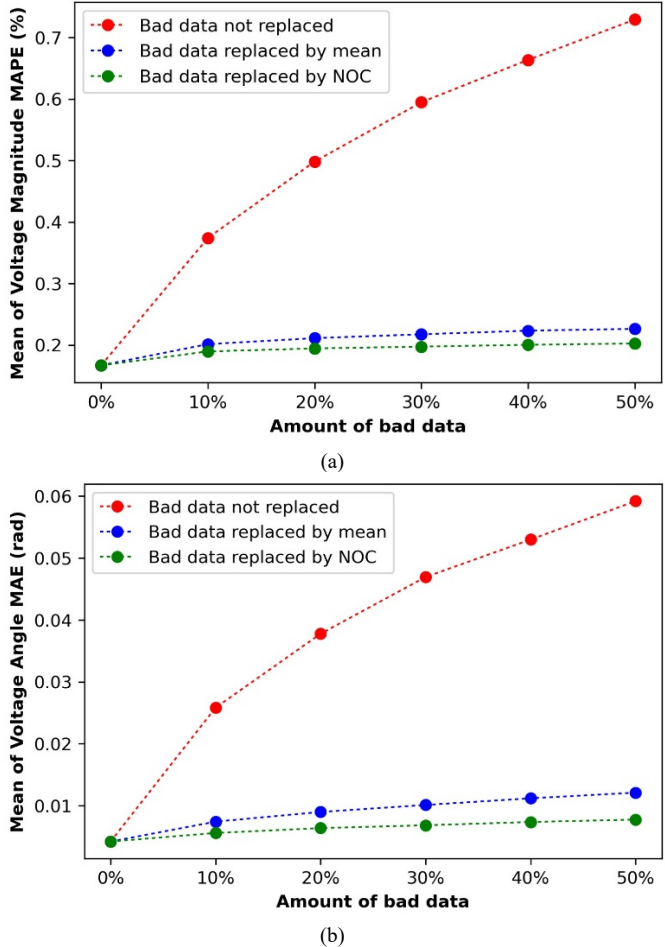

Figure 6: Bad data replacement with increasing probability of bad data

In the second scenario, we increase the severity of the bad data, while fixing the amount of testing samples that are bad. To do this, the severity was increased from $\sigma = 3\sigma_0$ to $\sigma = 7\sigma_0$, while setting $\eta = 30\%$. The results that were obtained when the proposed methodology is compared with the two cases described above (namely, no-replacement and replacement-by-mean), are shown in Figure 7. It is clear from the figure that the proposed methodology for correcting bad data (green dotted line) performs much better than the no-replacement case (red dotted line) and slightly better than the replacement-by-mean case (blue dotted line). Lastly, note that these studies were conducted on the trained DNN created in Section IV.A; i.e., only the inputs to the DNN in the testing phase were changed while its architecture was left unaltered.

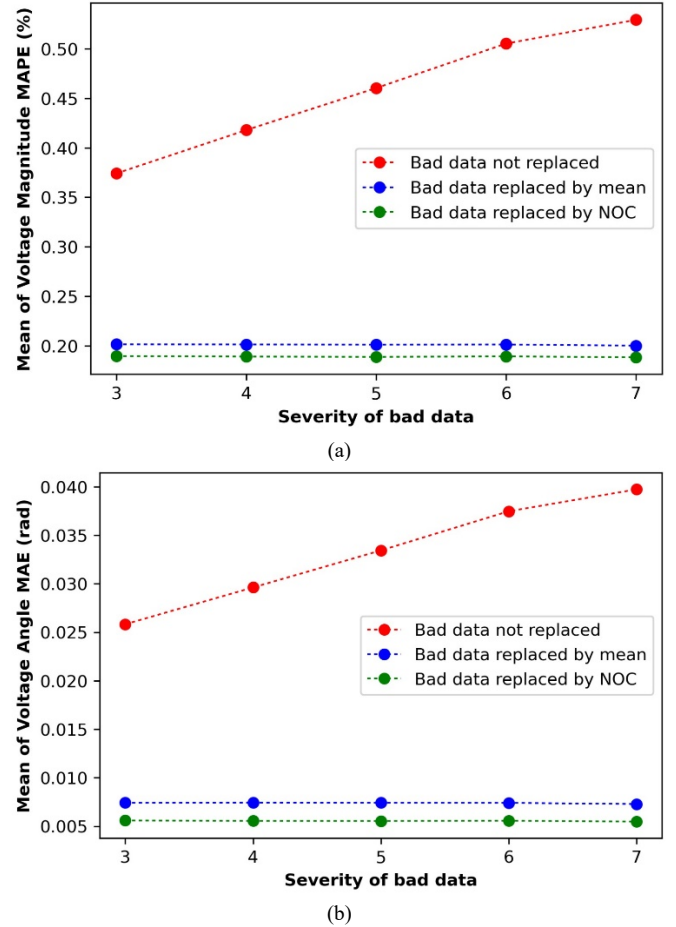

Figure 7: Bad data replacement with increasing severity of bad data

Considering the high-speed at which DeNSE is expected to operate during its online implementation (30 samples/second), it must be ensured that the Wald test and data preprocessing are performed within that time-frame. The most time-consuming portion in this regard is the proposed bad data correction module which must compare the current testing sample with all the samples in the training database to find the optimal replacement(s). It was observed that with 10,000 training samples and 41 phasor measurements as inputs, the bad data replacement for the 118-bus system could be carried out in 7.74±0.35 ms. As this is much less than the speed at which a PMU produces an output (≈33 ms), the proposed approach meets the high speed and high accuracy expectations of purely PMU-based state estimation.

*F. Tackling extreme scenarios*

In Section IV.E, the superiority of the BDDC methodology based on the Wald test and NOC was demonstrated. In this sub-section, the need and impact of the extreme scenario filter is discussed. 1,000 extreme scenarios were created for the 118-bus system by significantly increasing the loading of buses 8 and 10. Due to the physics of the power system, PMUs located on and in vicinity of 8 and 10, were impacted in these scenarios. Consequently, one or more measurements coming from the impacted PMUs (i.e., input features of the DeNSE) were flagged as bad data by the Wald test. At the same time, bad data was also added to the PMUs placed on buses 68 and 81 which are far away from the stressed region of the system. The extreme scenario filter identifies the set of features for which the BDDC should be suppressed, using the logic described in Section III.B. Three different outcomes were analyzed as shown in Table VII. Note that to obtain the results shown in this table, Gaussian noise was added to all the measurements.

The first row of Table VII depicts the outcome that was obtained when bad data was not corrected. The considerable deterioration of the results (compare this row with the first row of Table IV) was due to the presence of bad data in the measurements coming from PMUs placed at buses 68 and 81. A large amount of variability was also observed across the 1,000 scenarios as captured by the high standard deviation (std.) values. The second row of Table VII depicts the outcome that was obtained when BDDC took place but the extreme scenario filter was absent. The relatively high errors in this case were due to the presence of extreme scenarios around buses 8 and 10, whose corresponding PMU measurements were unnecessarily replaced. The best outcome was obtained when the proposed BDDC was applied to the PMU measurements coming from buses 68 and 81, but was suppressed by the extreme scenario filter for the PMU measurements coming from the region around buses 8 and 10, as depicted in the third row of Table VII. Thus, this analysis demonstrates the robust performance of the proposed DeNSE framework under diverse OCs.

TABLE VII
DeNSE PERFORMANCE WHEN BAD DATA AND EXTREME SCENARIO MANIFEST SIMULTANEOUSLY IN IEEE 118-BUS SYSTEM

| Method | Ave. Mag. MAPE (%) | | Ave. Angle MAE (rad) | |
|---|---|---|---|---|
| | Mean | Std. | Mean | Std. |
| DeNSE without BDDC | 0.3337 | 0.0254 | 0.0267 | 0.0023 |
| DeNSE with BDDC but without extreme scenario filter | 0.1853 | 0.0035 | 0.0059 | 0.0002 |
| DeNSE with BDDC and extreme scenario filter | 0.1812 | 0.0037 | 0.0053 | 0.0002 |

*G. Investigating impact of different database sizes*

In the proposed DeNSE, solving a variety of power flows under different operating conditions is necessary to create a comprehensive database for DNN training. The determination of the requisite number of samples is contingent upon the accuracy of the DNN relative to the number of samples utilized. In general, augmenting the training samples can further diminish DNN error until a point of performance saturation is reached. We realized this for the 118-bus system by progressively training our DNN with an increasing number of samples. We observed that beyond the threshold of 10,000 samples, no discernible improvement occurred (see Figure 8). Hence, we concluded that 10,000 samples (= sum of number of training and validation samples in Table III) was sufficient for robust performance of the DeNSE for this system.

*H. State estimation results for 2000-bus Texas system*

To demonstrate the applicability of the DeNSE to large transmission systems, we use the publicly available 2000-bus Synthetic Texas system [41]-[42]. The highest voltage buses in this system were 120 in number, and it was assumed that PMUs were already placed on these buses such that the voltage phasors of these buses as well as the current phasors of the lines that were coming out of these buses were measured by PMUs. By employing the time-series data available online for this system, the training and testing data was generated and a DNN was trained using the DeNSE framework explained in Section III.C.

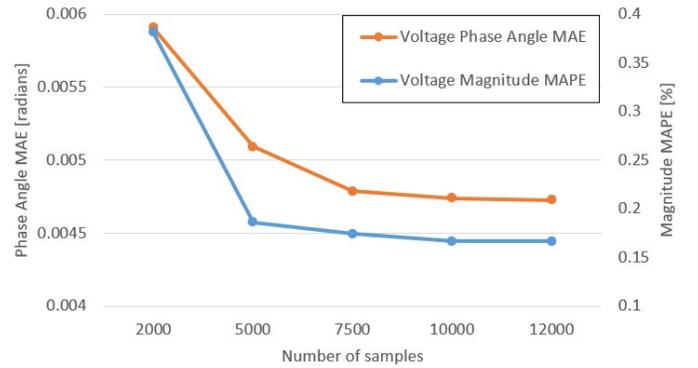

Figure 8: Impact of database sizes on DNN performance

The error estimates obtained with PMUs placed at 120 buses and under different noise models are shown in Table VIII and Figure 9, respectively. The outcomes presented in Table VIII correspond to a TVE of 1%, which is equivalent to a signal-to-noise ratio lying between 52 dB to 49 dB for Gaussian noise, 85 dB to 47 dB for GMM noise, and 90 dB to 85 dB for Laplacian noise models. Note that LSE for this system requires placing PMUs at 512 optimally selected buses. It can be observed from the table that with PMUs placed in less than one-quarter of the buses (120/512 = 0.234), the DeNSE has similar performance as LSE even in presence of non-Gaussian noise in PMU measurements. From Figure 9, it can be realized that the deterioration in the estimation performance is small even for buses that are 8 to 10 hops away. The hyperparameters obtained for the DeNSE for this system are summarized in Table IX. Note that the trained DNN took only 2.6 ms on average to produce the state estimates. This validates the ability of the DeNSE to estimate the states of large systems at high speeds.

TABLE VIII
PERFORMANCE OF DeNSE AND LSE UNDER DIFFERENT NOISE MODELS FOR 2000-BUS TEXAS SYSTEM

| Method (Noise Model) | Ave. Mag. MAPE (%) | Ave. Angle MAE (rad) | #Buses with PMUs |
|---|---|---|---|
| LSE (Gaussian) | 0.2809 | 0.0026 | 512* |
| DeNSE (Gaussian) | 0.2800 | 0.0024 | 120 |
| DeNSE (GMM) | 0.2714 | 0.0024 | 120 |
| DeNSE (Laplacian) | 0.2890 | 0.0027 | 120 |

* Optimally placed to ensure complete system observability

*Remark 1:* Note that for the test systems analyzed in this paper, the DeNSE performs state estimation using (1) in real-time based on a limited set of PMU measurements and without requiring knowledge of the system model and parameters. However, in the offline training phase, essential information is derived from power flow computations, which require knowledge of the system model and parameters. In other words, the proposed method remains model-agnostic during online operation but depends on system model and parameters during offline training. One way to avoid this dependency for an actual power system implementation is by directly utilizing historical SCADA state estimator results for creating the requisite training database of the DNN.

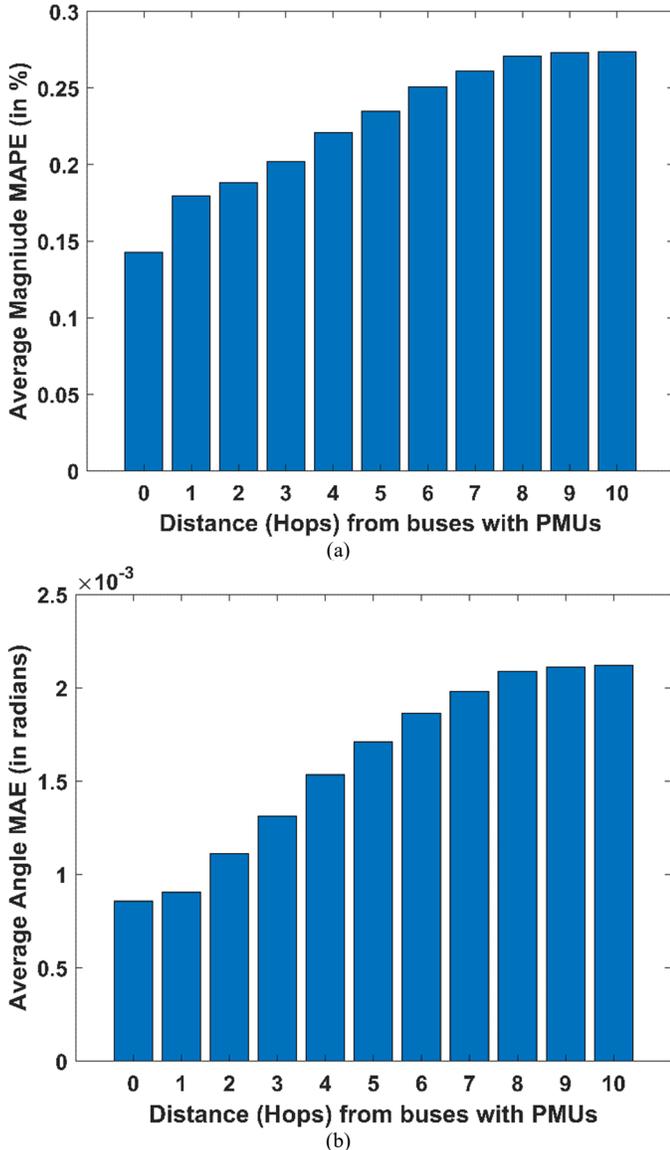

Figure 9: Performance evaluation of the DeNSE for the 2000-bus Synthetic Texas system as a function of the distance from the PMU-placed buses

*Remark 2:* When making additions to any existing system, a variety of factors must be considered. Therefore, it is not surprising that the final locations where PMUs would be placed are often decided based on negotiations with the grid operators, rather than through a purely mathematical optimization procedure (such as solving an OPP problem) [47]-[48]. However, this decision (of where to place the PMUs) does not affect the proposed approach because the locations of the PMUs are an input to the DeNSE, and not determined by the DeNSE. This means that the proposed approach is not limited to PMUs being placed only on the highest voltage buses of the system. In other words, even for a power system that has PMUs placed at low voltage buses, the DeNSE will simply take all available PMU measurements into consideration during training to give the state estimates of all the buses of that system during testing.

TABLE IX
HYPERPARAMETERS FOR DENSE FOR 2000-BUS TEXAS SYSTEM

| Hyperparameter | Value |
|---|---|
| Number of Hidden Layers | 4 |
| Number of Neurons per Hidden Layer | 500 |
| Activation Functions | ReLU (Hidden Layers) Linear (Output Layer) |
| Loss Function | Mean Squared Error |
| Optimizer | ADAM |
| Batch Size | 256 |
| Learning Rate | 0.001 |
| Number of Epochs | 3,000 |
| Early Stopping | Patience = 10 |
| Dropout | 30% |
| **Dataset size** | |
| Training | 7,500 |
| Validation | 2,500 |
| Testing | 4,000 |
| Total | 14,000 |

## V. CONCLUSION

In this paper, a Bayesian framework for high-speed time-synchronized TSSE was proposed that does not require complete observability of the system by PMUs for its successful execution. The proposed state estimator, called the DeNSE, overcame unobservability by indirectly combining inferences drawn from slow timescale SCADA data with fast timescale PMU measurements. The robustness of the proposed approach was demonstrated by its ability to successfully tackle practical challenges such as topology changes, non-Gaussian measurement noise, and different types of bad data under diverse operating conditions.

The IEEE 118-bus system and the 2000-bus Synthetic Texas system were used as the test systems for the analysis conducted here. In comparison to conventional approaches, the proposed DeNSE was able to bring the estimation errors of all the buses to reasonable levels while needing *less than half the number of PMUs* required for full observability for the 118-bus system and *less than one-quarter* for the 2000-bus system. The future scope of this work will involve developing strategies to further improve accuracy of the DeNSE by determining locations for adding new PMUs, extending the proposed method to handle events such as faults and load/generation losses, and providing provable performance guarantees [49].

## APPENDIX

### A. Logical explanation of DeNSE functioning

The DeNSE is an MMSE estimator in which the DNN approximates the conditional expectation, $\mathbb{E}(\boldsymbol{x}|\boldsymbol{z})$. For the $i^{th}$ state, $x_i$, the conditional expectation, $\mathbb{E}(x_i|\boldsymbol{z})$, can be written in terms of the probability distributions as shown below,

$$\mathbb{E}(x_i|\mathbf{z}) = \int_{-\infty}^{+\infty} x_i p(x_i|\mathbf{z}) dx_i = \int_{-\infty}^{+\infty} x_i \frac{p(x_i, \mathbf{z})}{p(\mathbf{z})} dx_i \quad (A.1)$$

where, $p(x_i|\mathbf{z})$ and $p(x_i, \mathbf{z})$ denote the conditional probability and the joint probability between $x_i$ and $\mathbf{z}$, respectively, and $p(\mathbf{z})$ denotes the probability distribution of $\mathbf{z}$. Now, it can be inferred from (1) and (A.1) that $\hat{x}_i(\mathbf{z})$ can be obtained for any value of $m$ (where $\mathbf{z} \in \mathbb{R}^m$), as long as one knows $p(x_i|\mathbf{z})$. Moreover, increasing $m$ can improve estimation quality only if the new measurements are *not* (a) correlated with the existing measurements, or (b) constant.

To better understand these inferences in the context of TSSE, consider the 3-bus system shown in Figure 10. The reference bus (bus 1) has an angle of 0°, but its magnitude is an unknown variable. Bus 2 has both load and generation, while bus 3 has only load. The system has three sensors (depicted by blue boxes) that are measuring the magnitude of the current flowing in line 1-2 and 2-1, and the magnitude of the current injection at bus 3. Let the goal be to estimate the voltage magnitude of bus 3 (i.e., $x_i = |V_3|$). The system is unobservable because $|V_3|$ cannot be estimated from the given measurements in the conventional least squares sense. Note that this example simply illustrates how the Bayesian framework of DeNSE can be used to estimate states which cannot be estimated using conventional methods due to limited observability. In an actual system the DeNSE will estimate *all* bus voltage magnitudes and angles without differentiating between unobserved buses, directly observed buses, and indirectly observed buses as it only relies on the joint PDF, $p(x_i, \mathbf{z})$, between the PMU measurements and the states.

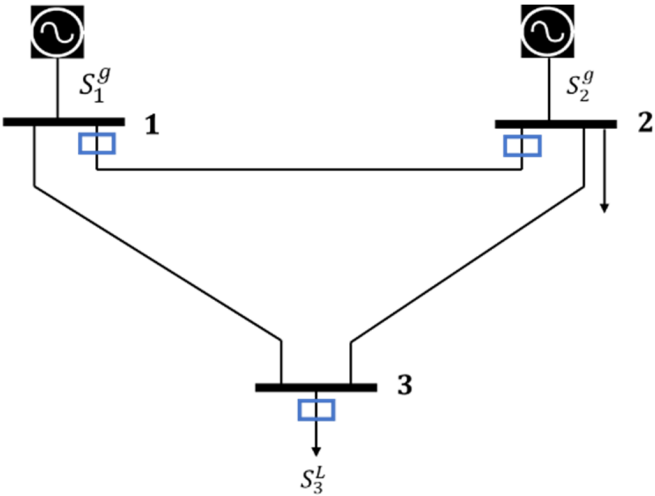

Figure 10: 3-bus system. Blue box indicates current magnitude sensor location

To generate $p(x_i, \mathbf{z})$ and $p(\mathbf{z})$ for this system, $F = 10{,}000$ power flows are solved. The parameters used for solving the power flows are provided in Table X. Due to the reasons mentioned in Section II.A, it is usually not possible to analytically compute $\mathbb{E}(x_i|\mathbf{z})$ for all $x_i$ and $\mathbf{z}$, which is why its approximation by a DNN is needed in the first place. However, for this 3-bus system, it was observed that the probability distributions of the relevant random variables ($|V_3|$, $|I_{12}|$, $|I_{21}|$, and $|I_3|$) could be well-approximated by multivariate normal distributions. In such a scenario, the conditional probability of $x_i$ given $\mathbf{z} = z_1, z_2, \ldots z_m$, can be written as [50],

$$p(x_i|\mathbf{z} = z_1, z_2, \ldots, z_m)$$
$$= \frac{1/\sqrt{(2\pi)^{m+1}|\Sigma_{y_p}|}\exp\left[-\frac{1}{2}(y_p - \mu_{y_p})^T \Sigma_{y_p}^{-1}(y_p - \mu_{y_p})\right]}{1/\sqrt{(2\pi)^m|\Sigma_{y_q}|}\exp\left[-\frac{1}{2}(y_q - \mu_{y_q})^T \Sigma_{y_q}^{-1}(y_q - \mu_{y_q})\right]} \quad (A.2)$$

where, $y_p$ and $y_q$ are obtained from power flow solutions with $y_p$ comprising all variables in $x_i$ and $\mathbf{z}$ and $y_q$ comprising variables in $\mathbf{z}$ only, $\mu$ and $\Sigma$ denote mean and covariance, and $|\Sigma|$ is determinant of the covariance. Now, using (A.1) and (A.2), we compare $\mathbb{E}(x_i|\mathbf{z})$ with the actual value of $x_i$ for five MMSE estimator cases: Case 1: $\mathbf{z} = \{\angle V_1\}$; Case 2: $\mathbf{z} = \{|I_{12}|\}$; Case 3: $\mathbf{z} = \{|I_3|\}$; Case 4: $\mathbf{z} = \{|I_{12}|, |I_{21}|\}$; Case 5: $\mathbf{z} = \{|I_{12}|, |I_3|\}$. Note that $|V_3|$, $|I_{12}|$, $|I_{21}|$, and $|I_3|$ are dependent variables as they correspond to converged power flow solutions, while $\angle V_1$ is a constant. The estimation results are shown in Table XI. In Case 1, $\mathbf{z}$ is a constant, and so $\mathbb{E}(x_i|\mathbf{z}) = \mathbb{E}(x_i)$, which is the mean value of $|V_3|$ across all $F$ power flows. As this case is not able to track the variations in operating conditions (OCs) across the different power flows, its estimate is the worst. Cases 2 and 3 gave similar results as they *separately* tracked the variations in $|I_{12}|$ and $|I_3|$ to estimate $|V_3|$. Despite having two measurements, the results of Case 4 are worse than both Case 2 and Case 3 because $|I_{12}|$ and $|I_{21}|$ are highly correlated. The expected values of Case 5 are closest to the ground-truth values as this estimator is able to use *both* $|I_{12}|$ and $|I_3|$ to estimate $|V_3|$. This analysis confirms that the knowledge of $p(\mathbf{x}|\mathbf{z})$ and not a large value of $m$, is the basis for the DeNSE to overcome unobservability. It is also worth mentioning that the estimation quality of the DeNSE improves if $F$ is increased, because with more training samples, the DNN will be able to better approximate the probability distributions, and in turn, $\mathbb{E}(\mathbf{x}|\mathbf{z})$.

TABLE X
SIMULATION PARAMETERS FOR 3-BUS SYSTEM IN P.U.

| Parameter | Values | Parameter | Values |
|---|---|---|---|
| Series Imp_1-2 | 0.05 + j0.1 | $P_2^g$ | $2 + N(0,0.04)$ |
| Series Imp_2-3 | 0 + j0.05 | $P_2^l$ | $0.5 + N(0,0.04)$ |
| Series Imp_3-1 | 0.02 + j0.05 | $Q_2^l$ | $0.1 + N(0,0.04)$ |
| Shunt Imp_1 | -j100 | $P_3^l$ | $2 + N(0,0.04)$ |
| Shunt Imp_2 | Inf | $Q_3^l$ | $0.5 + N(0,0.04)$ |
| Shunt Imp_3 | -j40 | $|V_1|$ | $1 + N(0,0.0001)$ |

TABLE XI
RESULTS OF CASE-STUDIES DONE ON 3-BUS SYSTEM

| Case | z | Mean Abs. Error = $\sum_{k=1}^{F}|x_i^k - \mathbb{E}(x_i^k|\mathbf{z}^k)|/F$ |
|---|---|---|
| 1 | $\{\angle V_1\}$ | 0.00100 |
| 2 | $\{|I_{12}|\}$ | 0.00014 |
| 3 | $\{|I_3|\}$ | 0.00021 |
| 4 | $\{|I_{12}|, |I_{21}|\}$ | 0.00094 |
| 5 | $\{|I_{12}|, |I_3|\}$ | 0.00005 |

B. *Python Resources for DeNSE Implementation*

All the Python codes required for implementing the DeNSE method developed in this paper can be accessed through the following GitHub repository: https://github.com/Anamitra-Pal-Lab/DeNSE. The Read Me file provided in this repository contains all the information that is needed to run the files and obtain the results.

REFERENCES

[1] G. Wang, G. B. Giannakis and J. Chen, "Robust and scalable power system state estimation via composite optimization," *IEEE Trans. Smart Grid*, vol. 10, no. 6, pp. 6137-6147, Nov. 2019.
[2] S. Chatzivasileiadis, *et al.*, "Micro-flexibility: Challenges for power system modeling and control," *Electric Power Syst. Research*, vol. 216, no. 109002, pp. 1-23, Mar. 2023.
[3] G. Wang, H. Zhu, G. B. Giannakis, and J. Sun, "Robust power system state estimation from rank-one measurements," *IEEE Trans. Control Network Syst.*, vol. 6, no. 4, pp. 1391-1403, Dec. 2019.
[4] A. S. Dobakhshari, M. Abdolmaleki, V. Terzija, and S. Azizi, "Robust hybrid linear state estimator utilizing SCADA and PMU measurements," *IEEE Trans. Power Syst.*, vol. 36, no. 2, pp. 1264-1273, Mar. 2021.
[5] M. Kabiri and N. Amjady, "A new hybrid state estimation considering different accuracy levels of PMU and SCADA measurements," *IEEE Trans. Instrum. Meas.*, vol. 68, no. 9, pp. 3078-3089, Sep. 2019.
[6] K. Sun, M. Huang, Z. Wei and G. Sun, "High-refresh-rate robust state estimation based on recursive correction for large-scale power systems," *IEEE Trans. Instrum. Meas.*, vol. 72, pp. 1-13, May 2023.
[7] J. Zhao, S. Wang, L. Mili, B. Amidan, R. Huang, and Z. Huang, "A robust state estimation framework considering measurement correlations and imperfect synchronization," *IEEE Trans. Power Syst.*, vol. 33, no. 4, pp. 4604-4613, Jul. 2018.
[8] P. Yang, Z. Tan, A. Wiesel, and A. Nehorai, "Power system state estimation using PMUs with imperfect synchronization," *IEEE Trans. Power Syst.*, vol. 28, no. 4, pp. 4162-4172, Nov. 2013.
[9] J. Zhao et al, "Power system real-time monitoring by using PMU-based robust state estimation method," *IEEE Trans. Smart Grid*, vol. 7, no. 1, pp. 300–309, Jan. 2016.
[10] N. M. Manousakis and G. N. Korres, "A hybrid power system state estimator using synchronized and unsynchronized sensors," *Int. Trans. Electr. Energ. Syst.*, vol. 38, no. 8, Aug. 2018.
[11] Z. Jin, P. Wall, Y. Chen, J. Yu, S. Chakrabarti and V. Terzija, "Analysis of hybrid state estimators: accuracy and convergence of estimator formulations," *IEEE Trans. Power Syst.*, vol. 34, no. 4, pp. 2565-2576, Jul. 2019.
[12] T. Chen, H. Ren, Y. Sun, M. Kraft, and G. A. J. Amaratunga, "Optimal placement of phasor measurement unit in smart grids considering multiple constraints," *J. Modern Power Syst. Clean Energy*, vol. 11, no. 2, pp. 479-488, Mar. 2023.
[13] A. Pal, G. A. Sanchez-Ayala, V. A. Centeno, and J. S. Thorp, "A PMU placement scheme ensuring real-time monitoring of critical buses of the network," *IEEE Trans. Power Del.*, vol. 29, no. 2, pp. 510-517, Apr. 2014.
[14] N. P. Theodorakatos, M. Lytras, and R. Babu, "Towards smart energy grids: a box-constrained nonlinear underdetermined model for power system observability using recursive quadratic programming," *Energies*, vol. 13, no. 7, pp. 1-17, Apr. 2020.
[15] M. A. R. S. Cruz, H. R. O. Rocha, M. H. M. Paiva, J. A. Lima Silva, E. Camby, M. E. V. Segatto, "PMU placement with multi-objective optimization considering resilient communication infrastructure," *Int. J. Electrical Power & Energy Syst.*, vol. 141, pp. 1-11, Apr. 2022.
[16] N. P. Theodorakatos, R. Babu, and A. P. Moschoudis, "The branch-and-bound algorithm in optimizing mathematical programming models to achieve power grid observability," *Axioms*, vol. 12, pp. 1-46, Nov. 2023.
[17] R. S. Biswas, B. Azimian, and A. Pal, "A micro-PMU placement scheme for distribution systems considering practical constraints," in *Proc. IEEE Power Eng. Soc. General Meeting*, Montreal, Canada, pp. 1-5, 2-6 Aug. 2020.
[18] K. Amare, V. A. Centeno, and A. Pal, "Unified PMU placement algorithm for power systems," in *Proc. IEEE North American Power Symposium (NAPS)*, Charlotte, NC, pp. 1-6, 4-6 Oct. 2015.
[19] M. Ghamsari-Yazdel, M. Esmaili, F. Aminifar, P. Gupta, A. Pal, and H. Shayanfar, "Incorporation of controlled islanding scenarios and complex substations in optimal WAMS design," *IEEE Trans. Power Syst.*, vol. 34, no. 5, pp. 3408-3416, Sep. 2019.
[20] U.S. Department of Energy, Office of Electricity Delivery and Energy Reliability, "Factors affecting PMU installation costs," Oct. 2014. [Online]. Available: https://www.smartgrid.gov/files/documents/PMU-cost-study-final-10162014.pdf
[21] A. Pal, C. Mishra, A. K. S. Vullikanti, and S. S. Ravi, "General optimal substation coverage algorithm for phasor measurement unit placement in practical systems," *IET Gener., Transm. Distrib.*, vol. 11, no. 2, pp. 347-353, Jan. 2017.
[22] A. Pal, A. K. S. Vullikanti and S. S. Ravi, "A PMU placement scheme considering realistic costs and modern trends in relaying," *IEEE Trans. Power Syst.*, vol. 32, no. 1, pp. 552-561, Jan. 2017.
[23] K. R. Mestav, J. Luengo-Rozas, and L. Tong, "Bayesian state estimation for unobservable distribution systems via deep learning," *IEEE Trans. Power Syst.*, vol. 34, no. 6, pp. 4910-4920, Nov. 2019.
[24] B. Azimian, R. S. Biswas, S. Moshtagh, A. Pal, L. Tong, and G. Dasarathy, "State and topology estimation for unobservable distribution systems using deep neural networks," *IEEE Trans. Instrum. Meas.*, vol. 71, pp. 1-14, Apr. 2022.
[25] K. R. Mestav and L. Tong, "Learning the unobservable: high-resolution state estimation via deep learning," in *Proc. 57th Annual Allerton Conf. Commun., Control, Comput.*, Monticello, IL, pp. 171-176, 24-27 Sep. 2019.
[26] G. Tian, Y. Gu, D. Shi, J. Fu, Z. Yu, and Q. Zhou, "Neural-network-based power system state estimation with extended observability," *J. Modern Power Syst. Clean Energy*, vol. 9, no. 5, pp. 1043-1053, Jun. 2021.
[27] V. Chakati, M. Pore, A. Pal, A. Banerjee, and S. K. S. Gupta, "Challenges and trade-offs of a cloud hosted phasor measurement unit-based linear state estimator," in *Proc. IEEE Power Eng. Soc. Conf. Innovative Smart Grid Technol.*, Washington DC, pp. 1-5, 23-26 Apr. 2017.
[28] R. Raz, "On the complexity of matrix product," in *Proc. 34th Annu. ACM Symp. Theory Computing*, pp. 144-151, 19 May 2002.
[29] E. Klarreich, "Multiplication hits the speed limit," *Commun. ACM*, vol. 63, no. 1, pp. 11-13, Jan. 2020.
[30] T. Ahmad and N. Senroy, "Statistical characterization of PMU error for robust WAMS based analytics," *IEEE Trans. Power Syst.*, vol. 35, no. 2, pp. 920-928, Mar. 2020.
[31] D. Salls, J. Ramirez, A. Varghese, J. Patterson, and A. Pal, "Statistical characterization of random errors present in synchrophasor measurements," in *Proc. IEEE Power Eng. Soc. General Meeting*, Washington DC, pp. 1-5, 26-29 Jul. 2021.
[32] A. C. Varghese, A. Pal, and G. Dasarathy, "Transmission line parameter estimation under non-Gaussian measurement noise," *IEEE Trans. Power Syst.*, vol. 38, no. 4, pp. 3147-3162, Jul. 2023.
[33] J. Zhao and L. Mili, "A framework for robust hybrid state estimation with unknown measurement noise statistics," *IEEE Trans. Industrial Informatics*, vol. 14, no. 5, pp. 1866-1875, May 2018.
[34] Y. Gu, Z. Yu, R. Diao and D. Shi, "Doubly-fed deep learning method for bad data identification in linear state estimation," *J. Modern Power Syst. Clean Energy*, vol. 8, no. 6, pp. 1140-1150, Nov. 2020.
[35] W. Liu, J. Liu, H. Li, Q. Du and Y. -L. Wang, "Multichannel signal detection based on Wald test in subspace interference and Gaussian noise," *IEEE Trans. Aerospace Electron. Syst.*, vol. 55, no. 3, pp. 1370-1381, Jun. 2019.
[36] Y. Yang, Z. Yang, J. Yu, K. Xie and L. Jin, "Fast economic dispatch in smart grids using deep learning: an active constraint screening approach," *IEEE Internet Things J.*, vol. 7, no. 11, pp. 11030-11040, Nov. 2020.
[37] J. A. D. Massignan, J. B. A. London and V. Miranda, "Tracking power system state evolution with maximum-correntropy-based extended Kalman filter," *J. Modern Power Syst. Clean Energy*, vol. 8, no. 4, pp. 616-626, Jul. 2020.
[38] S. J. Pan, and Q. Yang, "A survey on transfer learning," *IEEE Trans. Knowledge Data Eng.*, vol. 22, no. 10, pp. 1345-1359, Oct. 2010.
[39] Y. Zhang, Y. Zhang, and Q. Yang, "Parameter transfer unit for deep neural networks," *Advances Knowledge Discovery Data Mining*, Cham: Springer International Publishing, pp. 82–95, Mar. 2019.
[40] K. D. Jones, A. Pal, and J. S. Thorp, "Methodology for performing synchrophasor data conditioning and validation," *IEEE Trans. Power Syst.*, vol. 30, no. 3, pp. 1121-1130, May 2015.
[41] H. Li, J. H. Yeo, A. L. Bornsheuer and T. J. Overbye, "The creation and validation of load time series for synthetic electric power systems," *IEEE Trans. Power Syst.*, vol. 36, no. 2, pp. 961-969, Mar. 2021.
[42] A. B. Birchfield, T. Xu, and T. J. Overbye, "Power flow convergence and reactive power planning in the creation of large synthetic grids," *IEEE Trans. Power Syst.*, vol. 33, no. 6, pp. 6667-6674, Nov. 2018.
[43] F. Chollet et al., "Keras," 2015. [Online]. Available: https://keras.io
[44] WANDB, "Weights & Biases", Accessed on Nov. 22, 2023. [Online]. Available: https://wandb.ai/site
[45] "IEEE/IEC International Standard - Measuring relays and protection equipment - Part 118-1: Synchrophasor for power systems - Measurements," *IEC/IEEE 60255-118-1:2018*, pp.1-78, Dec. 2018.
[46] Q. Yang, A. Sadeghi and G. Wang, "Data-driven priors for robust PSSE via Gauss-Newton unrolled neural networks," *IEEE J. Emerging Selected Topics Circuits Syst.*, vol. 12, no. 1, pp. 172-181, Mar. 2022.

[47] D. Borkowski, A. Wetula, J. Kowalski, S. Barczentewicz, J. Nabielec, M. Rogóż, and I. Szczygieł, "Experimental setup for harmonic impedance measurement in a real HV power grid," *Electr. Power Comp. Syst.*, vol. 47, no. 8, pp. 733-742, Aug. 2019.
[48] N. Theodorakatos, N. Manousakis, and G. Korres, "Optimal placement of phasor measurement units with linear and non-linear models," *Electr. Power Comp. Syst.*, vol. 43, no. 4, pp. 357–373, Feb. 2015.
[49] B. Azimian, S. Moshtagh, A. Pal, and S. Ma, "Analytical verification of deep neural network performance for time-synchronized distribution system state estimation," *J. Modern Power Syst. Clean Energy* (Early Access).
[50] H. Pishro-Nik, "Introduction to probability, statistics, and random processes", *Kappa Research LLC*, 2014. [Online]. Available: https://www.probabilitycourse.com

**Antos Cheeramban Varghese** earned his B. Tech degree in Electrical Engineering from the Indian Institute of Technology Jodhpur, Rajasthan, India, in 2016. He subsequently completed his M. Tech degree in Electrical Engineering at the Indian Institute of Technology Bombay, Mumbai, India, in 2018. Currently, he is pursuing his Ph.D. degree in Electrical Engineering at Arizona State University in Tempe. His research is focused on leveraging machine learning, statistical analysis, and optimization techniques to address various challenges within the field of power systems.

**Hritik Gopal Shah** received the B.E. degree in Electronics Engineering from the DJ Sanghvi College of Engineering, Mumbai, India, in 2021 and the M.S. in Industrial Engineering from Arizona State University, Tempe, AZ, USA, in 2023. He is a Data Science Analyst with Reliability and Resiliency Group at Eversource Energy, Berlin, CT. He is working on projects related to Power Grid Reliability, Outage Management and Advanced Load Forecasting.

**Behrouz Azimian** (Student Member, IEEE) received the Bachelor of Science degree in electrical engineering from Iran University of Science and Technology, Tehran, Iran, in 2016. He Received the Master of Science degree in electrical engineering from Alfred University, NY, USA, in 2019. He is now a Ph.D. student at Arizona State University. His research interests include machine learning, deep learning, application of artificial intelligence and optimization methods in power system problems such as state estimation, electricity markets, and electric vehicle charging scheduling.

**Anamitra Pal** (Senior Member, IEEE) received the B.E. degree (summa cum laude) in electrical and electronics engineering from the Birla Institute of Technology at Mesra, Ranchi, India, in 2008, and the M.S. and Ph.D. degrees in electrical engineering from Virginia Tech, Blacksburg, VA, USA, in 2012 and 2014, respectively. He is currently an Associate Professor in the School of Electrical, Computer, and Energy Engineering at Arizona State University (ASU), Tempe, AZ, USA. His research interests include data analytics with a special emphasis on time-synchronized measurements, artificial intelligence-applications in power systems, renewable generation integration studies, and critical infrastructure resilience. Dr. Pal has received the 2018 Young CRITIS Award for his contributions to the field of critical infrastructure protection, the 2019 Outstanding Young Professional Award from the IEEE Phoenix Section, the National Science Foundation CAREER Award in 2022, and the 2023 Centennial Professorship Award from ASU.

**Evangelos Farantatos** received the Diploma in Electrical and Computer Engineering from the National Technical University of Athens, Greece, in 2006 and the M.S. and Ph.D. degrees from the Georgia Institute of Technology, Atlanta, GA, USA, in 2009 and 2012, respectively. He is a Sr. Principal Team Lead with the Transmission Operations and Planning R&D Group at EPRI, Palo Alto, CA. He is managing and leading the technical work of various R&D projects related to synchrophasor technology, power systems monitoring and control, power systems stability and dynamics, renewable energy resources modeling, grid operation protection and control with high levels of inverter-based resources. He is a Senior Member of IEEE.